\documentstyle[sprocl]{article}
\input epsf.tex
\def\DESepsf(#1 width #2){ \epsfxsize = #2 \epsfbox{#1}}
\def\be{\begin{equation}}
\def\ee{\end{equation}}
\def\bea{\begin{eqnarray}}
\def\eea{\end{eqnarray}}

\begin{document}
\title{ 
Physics with $B \to \eta'$ and $B\to J/\psi$
~\footnote{Talk presented 
at the 4th International Workshop on Particle Physics 
Phenomenology, Kaohsiung, Taiwan, China, 18-21 June, 
1998}}

\vfill
\author{Xiao-Gang He}
\address{
\rm  Department of Physics, National Taiwan University,
Taipei, 10764, Taiwan
}
\maketitle\abstracts{
In this talk I discuss two topics related to B decays. I will first
discuss the Standard Model prediction for $B\to \eta' X_s$, and then 
discuss how to use experimental information from $B\to J/\psi K^*$
to test CP violation beyond the SM.
}

I will report some of my recent work on $B\to \eta' X_s$
 and
$B\to J/\psi K^*$ with my collaborators~\cite{hg3,he4,he5} in this talk. 
The processes $B\to \eta' X_s$ and $B\to J/\psi K^*$ 
have been analyzed in detail by the CLEO collaboration~\cite{CLEO2,Pwave}. 
The unexpected large branching ratio of~\cite{CLEO2}
 $(6.2\pm 1.6\pm 1.3)\times 
10^{-4}$ for 
$B\to \eta' X_s$ has led some to claim that new physics is needed 
to explain the observation by CLEO~\cite{HT,KP,FR}. 
I will show that in fact the 
Standard Model prediction is consistent with experimental data when 
related processes are treated properly~\cite{hg3}. The measurement of the 
full angular analysis of~\cite{Pwave} $B\to J/\psi K^*$ from CLEO 
has also led to some interesting
physics. I will show that information obtained from the angular distribution
for this decay provide good test for CP violation beyond the SM~\cite{he4,he5}.

\section{ $B\to \eta' X_s$ in the Standard Model}

The recent observation of $B\to \eta' X_s$~\cite{CLEO2} and 
$B\to \eta^{\prime}K$~\cite{CLEO1} decays with high momentum
$\eta^{\prime}$ has stimulated many theoretical 
activities~\cite{HT,KP,FR,AS,he1,excl,excl1,HZ}. 
One of the mechanisms proposed to account for this decay is
$b\to sg^*\to
sg\eta^{\prime}$~\cite{HT,AS} where the $\eta^{\prime}$ meson is  
produced 
via the anomalous 
$\eta'-g-g$ coupling. According to a previous analysis~\cite{HT},
this mechanism within the Standard Model(SM) can only account for 1/3 
of the 
measured branching ratio~\cite{CLEO2}. 
There are also other calculations of $B \to \eta' X_s$ 
based on four-quark operators of the effective  
weak-Hamiltonian~\cite{KP,he1}. 
These contributions to the branching ratio, typically $10^{-4}$, 
are also too small to account for $B\to \eta' X_s$, 
although the four-quark-operator
contribution is capable of explaining the branching ratio for
the exclusive $B \to \eta' K$ decays~\cite{excl,excl1}.
These results
have inspired proposals for an enhanced $b\to sg$ and other 
mechanisms arising from physics 
beyond the SM~\cite{HT,KP,FR}. 
It will be shown here that the SM is in fact consistent with 
experimental data from CLEO~\cite{hg3}.

The quark level effective Hamiltonian for the $B\to \eta'  
X_s$ decay is 
given by~\cite{REVIEW}:
\begin{eqnarray}
H_{eff}(\Delta B=1)&=&{G_F\over  
\sqrt{2}}[\sum_{f=u,c}V_{fb}V_{fs}^*(C_1(\mu)O_1^f(\mu)+C_2(\mu)O_2^f(\mu))
\nonumber\\
&-&V^*_{ts}V_{tb}\sum_{i=3}^{6}(C_i^{eff}(\mu)O_i(\mu)
+C_8^{eff}(\mu)O_8(\mu))],
\label{HAMI}
\end{eqnarray}
The operators are defined in Ref.[15]. $C_i$ are the Wilson Coefficients (WC).
The values for the WC's given in Ref.[16,17] will be used in this discussion. 
The superscript $^{eff}$ indicate 
that the matrix element corrections are also included.

Let me first  discuss 
the four-quark operator contributions to 
$B\to \eta' X_s$. 
The four-quark operators
can induce three types of processes represented by
1) $<\eta'|\bar q \Gamma_1 b|B> <X_s|\bar s \Gamma_1'q|0>$,
2) $<\eta'|\bar q \Gamma_2 q|0><X_s|\bar s \Gamma b| B>$, and 
3) $<\eta' X_s|\bar s \Gamma_3q|0><0|\bar q\Gamma_3'|B>$. 
Here $\Gamma^{(')}_i$ denotes appropriate gamma matrices. 
The contribution from 1) gives a ``three-body'' type of decay,
$B\to \eta' s \bar q$. The contribution from 2) gives a ``two-body''
type of decay $b\to s\eta'$.
And the contribution from 3) is the annihilation
type which is relatively suppressed and will be neglected.
Several decay constants and form factors needed in the calculations  
are
listed below:
\begin{eqnarray}
&&<0|\bar u\gamma_\mu \gamma_5 u|\eta'> = 
<0|\bar d \gamma_\mu \gamma_5 d|\eta'>
=if_{\eta'}^u p^{\eta'}_\mu\nonumber\\
&&<0|\bar s \gamma_\mu \gamma_5 s|\eta'> = 
if_{\eta'}^s p^{\eta'}_\mu,\;\;
<0|\bar s \gamma_5 s|\eta'> = i(f_{\eta'}^u-f_{\eta'}^s) 
{m^2_{\eta'}\over 2m_s},\nonumber\\
&&f_{\eta'}^u = {1\over \sqrt{3}
} (f_1 \cos\theta_1 + 
{1\over \sqrt{2}} f_8 \sin\theta_8),\;\;
f_{\eta'}^s = {1\over \sqrt{3}}
(f_1\cos\theta_1 - \sqrt{2} f_8 \sin\theta_8),\nonumber\\
&&<\eta'|\bar q\gamma_\mu b|B>
= F_1^{Bq}(p^B_\mu + p^{\eta'}_\mu) +
(F_0^{Bq}-F_1^{Bq}) {mB^2-m_{\eta'}^2\over 
q^2} q_\mu,\nonumber\\
&&F_{1,0}^{Bq}={1\over \sqrt{3}} ({1\over \sqrt{2}} 
\sin\theta F^{B\eta_8}_{1,0}
+\cos\theta F^{B\eta_1}_{1,0}).
\end{eqnarray}
For the $\eta'-\eta$ mixing associated with decay constants above,
I have used the two-angle 
parameterization.
The numerical values of various parameters are obtained from 
Ref. [18] with 
$f_1= 157$ MeV, $f_8=168$ MeV,
and the mixing angles $\theta_1 = -9.1^0$, $\theta_8=-22.1^0$. 
For the mixing angle associated with form factors, I used the
one-angle parameterization with\cite{FK} $\theta = - 15.4^o$,
since these form factors were calculated in that 
formulation~\cite{he1,KP}. 
In the latter discussion of $b\to sg\eta'$,  
I shall use the same parameterization in order to compare our results with  
those of earlier works
\cite{AS,HT}.  
For form factors, I assume
that $F^{B\eta_1} = F^{B\eta_8} = F^{B\pi}$ with 
dipole and monopole $q^2$ dependence for $F_1$ and $F_0$,  
respectively.
I used the running mass $m_s \approx 120$ MeV
at $\mu = 2.5$ GeV and $F^{B\pi} = 0.33$ following Ref.[13].

Using $V_{ts} = 0.038$, $\gamma = 64^0$ and $\mu = 5$ GeV, it is found  
that
the branching ratios in the signal region $p_{\eta'} > 2.2$ GeV 
($m_X < 2.35$ GeV) are given by
\begin{eqnarray}
B(b\to \eta' s) = 0.9\times 10^{-4},\;\;
B(B\to \eta' s \bar q) = 0.1\times 10^{-4}.
\end{eqnarray}
The branching ratio can reach $2\times 10^{-4}$ if all parameters  
take 
values in favor of $B\to \eta' X_s$.
Clearly the mechanism by four-quark operator is not sufficient  
to explain the observed $B\to \eta^{\prime}X_s$ 
branching ratio.

I now turn to 
$b\to \eta' s g$ through the QCD anomaly. To see how the
effective Hamiltonian in Eq. (\ref{HAMI}) can be applied to calculate
this process,  
let me  rearrange the effective Hamiltonian such that
\begin{eqnarray}
&&\sum_{i=3}^6C_iO_i=(C_3^{eff}+{C_4^{eff}\over N_c})O_3
+(C_5^{eff}+{C_6^{eff}\over N_c})O_5\nonumber\\
&&
- 2(C_4^{eff}-C_6^{eff})O_A+2(C_4^{eff}+C_6^{eff})O_V+C_8^{eff}O_8,
\label{GLUE}
\end{eqnarray}
where 
\begin{equation}
O_A=\bar{s}\gamma_{\mu}(1-\gamma_5)T^a b \sum_{q}\bar{q}\gamma^{\mu}
\gamma_5T^a q,
\;\;O_V=\bar{s}\gamma_{\mu}(1-\gamma_5)T^a b \sum_{q}
\bar{q}\gamma^{\mu}
T^a q.
\end{equation} 
Since the light-quark bilinear in $O_V$ carries the quantum number
of a gluon, one expects~\cite{AS} $O_V$ give contribution to
the $b\to sg^*$ form factors. In fact, by applying the QCD equation  
of motion
: $D_{\nu}G^{\mu\nu}_a=g_s\sum \bar{q}\gamma^{\mu}T^a q$,
one obtains, $O_V=(1/ g_s)\bar{s}\gamma_{\mu}(1-\gamma_5)T^a b D_{\nu}
G^{\mu\nu}_a$.   
The effective $b\to sg^*$ vertex can be written as
\begin{eqnarray}
\Gamma_{\mu}^{bsg}=-{G_F\over \sqrt{2}}  V_{ts}^*V_{tb}
{g_s\over 4\pi^2} (\Delta F_1 \bar s( q^2 \gamma_\mu - q\!\!\!/\  
q_\mu) LT^a b - i F_2 m_b \bar s \sigma_{\mu\nu}q^\nu RT^ab) .
\label{DECOM}
\end{eqnarray}
The form
factors $\Delta F_1$ and $F_2$ are defined according to the convention in 
Ref. [6].  One obtains 
\begin{eqnarray}
&&\Delta F_1 ={4\pi \over \alpha_s} (C_4^{eff}(\mu)+C_6^{eff}(\mu)),\;\;
F_2 =-2C_8^{eff}(\mu)
\label{F12}
\end{eqnarray}
Note that the relative sign of $\Delta F_1$ and $F_2$ obtained agree with 
those in Refs. [6,7] and [19] which
results in a destructive interference.

At the hadronic level
the anomalous $\eta'-g-g$ coupling is given by: 
$a_g(\mu) \cos\theta \epsilon_{\mu\nu\alpha\beta}q^\alpha k^\beta$ with
$a_g(\mu) = \sqrt{N_F}
\alpha_s(\mu)/\pi f_{\eta'}$, q and k the momenta of the two gluons.
Using the WC's at $\mu=5$ GeV, 
the branching ratio is found to be $B(b\to sg\eta')=5.6\times  
10^{-4}$
with a cut on $m_{X}\equiv \sqrt{(k+p')^2}\leq 2.35$ GeV. 
The spectrum for         
$dB(b\to sg\eta')/dm_X$ is depicted in Fig. 1. 
The peak of  
the 
spectrum corresponds to $m_X\approx 2.4$ GeV. 
The destructive interference between $F_1$ and $F_2$ lowers down the 
branching ratio by about 14\% which is quite different from the results 
obtained in Refs.[6,10] because our $\Delta F_1$ is larger than
theirs because the inclusion of the matrix element corrections.

\begin{figure}[htb]
\centerline{\DESepsf(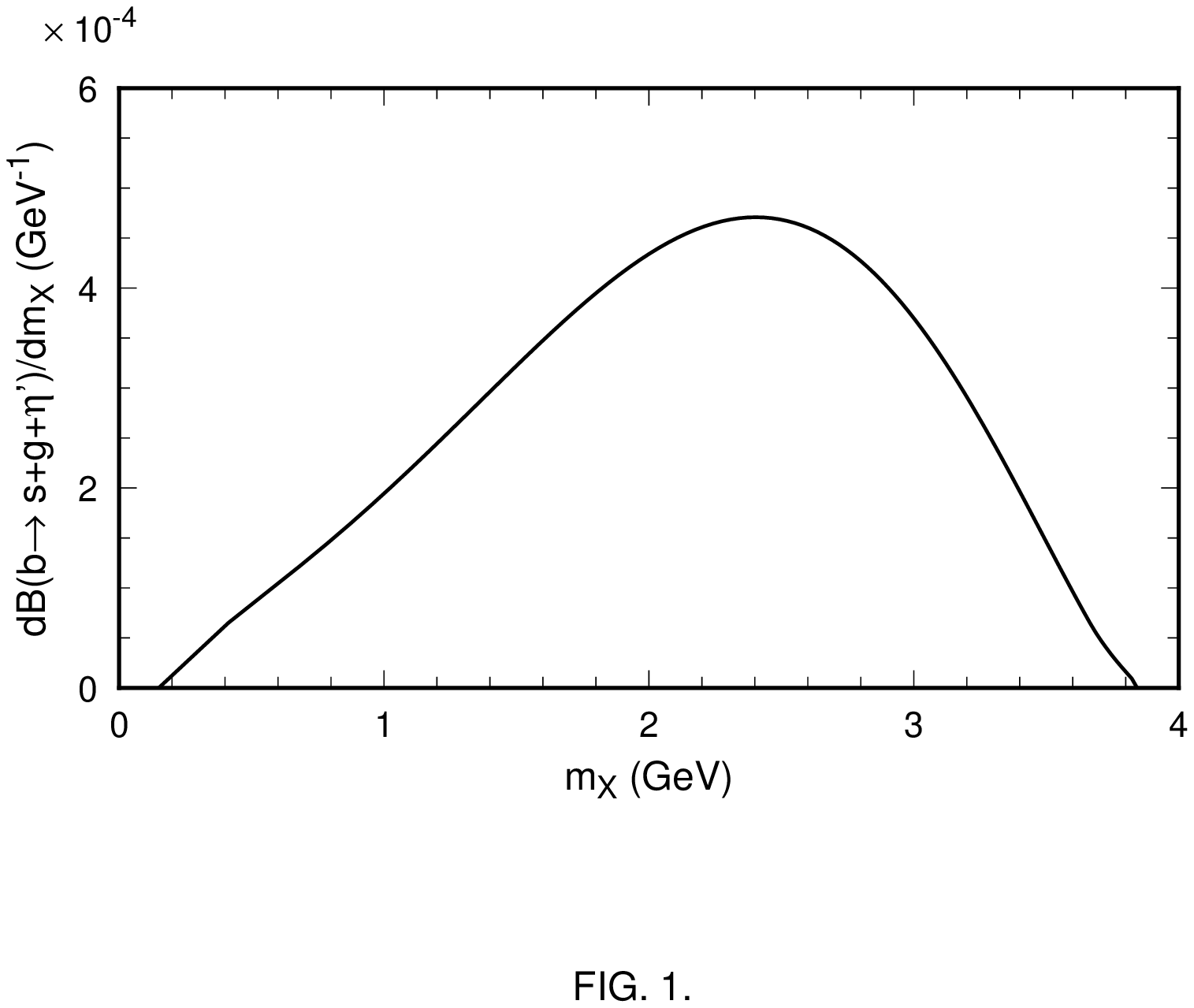 width 6 cm)} 
\caption{
The distribution of $B(b\to s+g+\eta')$
as a function of the recoil mass 
$m_X$.
}
\end{figure}

In the above calculation, $a_g(\mu)$ of the  
$\eta'-g-g$ vertex is treated as a constant independent 
of invariant-masses of the gluons, and $\mu$ is 
set to be $5$ GeV. In practice, $a_g(\mu)$ should behave like 
a form-factor which becomes suppressed as the gluons attached to it 
go farther off-shell~\cite{KP}. 
It is possible that the branching ratio 
obtained gets reduced significantly by the 
form-factor effect in $\eta'-g-g$ vertex.
Should a large form-factor suppression occur, the additional 
contribution from $b\to \eta' s$ and $B\to \eta' s \bar q$ 
discussed earlier will become crucial. 
I however would like to stress that 
the estimate of $b\to sg\eta^{\prime}$
with $\alpha_s$ evaluated at $\mu=5$ GeV is conservative. To  
illustrate 
this, let me compare branching ratios for $b\to sg\eta^{\prime}$ 
obtained at $\mu=5$ GeV
and $\mu=2.5$ GeV respectively. 
The branching ratios at the 
above two scales with the kinematic al cut on $m_X$ are $4.9\times 10^{-4}$ and 
$8.5\times 10^{-4}$ respectively. One can clearly see the significant
scale-dependence! With the enhancement resulting from lowering the 
renormalization scale, there seems to be some room for the  
form-factor 
suppression in the attempt of explaining $B\to \eta^{\prime}X_s$ by
$b\to sg\eta^{\prime}$.

It is clear that the Standard Model prediction for $B\to \eta' X_s$ is
not in conflict with experimental data from CLEO.

\section{$B\to J/\psi K^*$ and Test for CP Violation Beyond SM}

The CLEO Collaboration has recently reported \cite{Pwave} 
the first full angular analysis of 
$B\to  J/\psi K^{*0}$ decays.
They find that the $P$ wave component is small,
$\vert P\vert^2=|A_T|^2 = 0.16 \pm0.08\pm 0.04$, and final state interaction 
(FSI) phases 
$\phi(A_T)= -0.11\pm0.46\pm 0.03\;rad$, and $\phi(A_{||})=3.00\pm 0.37\pm 
0.04\;rad$ which are consistent with zero or $\pi$ FSI phases in the 
convention where $\phi(A_0) = 0$.
The small value for $|P|^2$ shows that 
$B\to  J/\psi K^{*0} \to J/\psi K_S \pi^0$ decay is
dominated by $CP$-even  final states.
This makes it practical to use $B\to J/\psi K^*
\to J/\psi K_S \pi^0$ to observing mixing induced CP violation and 
to measure $\sin 2\beta$ without
invoking an angular analysis \cite{DQSTL}.
The difference between
$\sin2\beta$ measured from $B\to J/\psi K_S$ ($\sin2\beta_{J/\psi K_s}$)
 and $B\to J/\psi K^*$ ($\sin2\beta_{J/\psi K_S \pi^0}$)
 can provide good test
for CP violation beyond the SM~\cite{he4}.
With increased luminosities at CLEO, B-factories and other 
facilities, the charges of the B aand $K^*$ can be identified, and therefore
it is 
also possible to study 
direct CP violation in $B\to J/\psi K^*$ decays~\cite{he5}. 

To proceed, let me show how 
$\sin 2\beta_{J/\psi K_S}$ and
$\sin 2\beta_{J/\psi K_S \pi^0}$ can be used to test CP violation beyond 
the SM.
The usual mixing induced $CP$ violation measure
is ${\rm Im\,}\xi = {\rm Im\,}\{(q/p)(A^*\bar A/\vert A\vert ^2)\}$,
where $q/p = e^{-2i\phi_B}$ is from $B^0$--$\bar B^0$ mixing,
while $A$, $\bar A$ are
$B$,  $\bar B$ decay amplitudes into the same final state.
For $B\rightarrow J/\psi K_S$, the final state is $P$-wave hence $CP$ odd.
Setting the weak phase to be $\sigma_S$, one has
\begin{eqnarray}
{\rm Im\,}\xi(B\rightarrow J/\psi K_S) = -\sin(2\phi_B + 2\sigma_S)
\equiv - \sin2\beta_{J/\psi K_S}.
\nonumber
\end{eqnarray}
For $B\rightarrow J/\psi K^*\rightarrow J/\psi K_S \pi^0$,
the final state has both $P$-amplitude ($CP$ odd) and
$S$- and $D$-amplitudes ($CP$ even) (or their linear combinations 
$A_{||}$ and $A_0$). Let me denote their corresponding 
weak phases by $\sigma_T$, and $\sigma_{||}$ and $\sigma_0$, 
respectively.
If $A_{||}$ and  $A_0$ amplitudes
 have a common weak phase $\sigma_*$,
one has
\begin{eqnarray}
{\rm Im\,}\xi(B\rightarrow J/\psi K_S \pi^0)
&=& {\rm Im\,} \{ e^{-2i\phi_B} [ e^{-2i\sigma_T}\vert P \vert ^2
              -e^{-2i\sigma_*}(1-\vert P \vert ^2)]\}\nonumber\\
&\equiv&- (1-2\vert P \vert ^2)\sin2\beta_{J/\psi K_S\pi^0},
%\nonumber
\end{eqnarray}
In the SM one obtains the usual result of
$\sin2\beta_{J/\psi K_S}= \sin2\beta_{J/\psi K_S\pi^0}= \sin2\beta$.
Clearly, both measurements provide true information about $\sin2\beta$
within SM.
However, this is no longer true if one goes beyond SM.
Even for the case where the weak phases of $A_{||}$ 
and $A_0$ amplitudes are
equal, if $\sigma_T\neq \sigma_*$ or if $\sigma_T,\sigma_* \neq \sigma_S$,
then $\sin2\beta_{J/\psi K_S} \neq \sin2\beta_{J/\psi K_S\pi^0}$ follows.

To the first order in new weak phases, one has

\begin{eqnarray}
\Delta_{KK^*}\approx (2\sigma_S -2\sigma_*+{2|P|^2\over 1-2|P|^2}(\sigma_T
-\sigma_*))\cos(2\phi_B).
\end{eqnarray}
If experiments will measure a non-zero $\Delta_{KK^*}$. It is a signal for 
CP violation beyond the SM.

There are many ways where new physics may change the phases $\sigma_i$. To the
lowest order they may arise from dimension 6 four quark operators and 
dimension 5 color dipole moment operators.
New physics contributions 
from $C_{L}
\bar c \gamma^\mu(1\pm \gamma_5) c \bar s \gamma_\mu (1-\gamma_5) b$
type of four quark 
interactions are proportional to the dominant SM contribution in Eq.[1] which 
just generate a common weak phase for all the amplitudes and therefore 
$\Delta_{KK^*}$  is zero like the SM prediction.
$\Delta_{KK^*}$ discussed here does not provide good test
for new physics of this type.
The interaction of the form
$C_{R}\bar c\gamma^\mu (1\pm \gamma_5) c \bar s \gamma_\mu (1+\gamma_5) b$
will generate different phases for $A_T$ and $A_{||,0}$.
The reason is that the current
$\bar s \gamma^\mu(1+\gamma_5) b$ contribution to $A_T$ is proportional to
$C_{SM}+ C_R$, but to $A_{||,0}$ is proportional to $C_{SM}-C_R$. 
To a good approximation,
$\delta_T = - \delta_{||,0}$. $\Delta_{KK^*}$ can be different
from zero. 

As an example, let us study R-parity violating 
supersymmetric models.
In R-parity violating supersymmetric models, exchange of charged sleptons or
down type squarks can generate non-zero $C_R$ with an arbitrary phase 
$\delta_R$. The allowed value for $C_R$ is constrained from 
experimental data on $b\to s\gamma$. This still allow $C_R$ term to 
contribute to $B\to J/\psi K^*$ at the amplitude level as large as
20\% of the SM contribution. Stronger constraints can be obtained
by assuming that $b\to c\bar c s$ is similar in strength to 
$b\to c\bar u s$~\cite{GW}. 
The upper bound of the weak phases are approximately given by

\begin{eqnarray}
\delta_T=-\delta_* \approx 0.1\sin \delta_R.
\end{eqnarray}

Using Eq.(9), 
one can easily see that 
the difference $\Delta_{KK^*}$ can reach $-0.35 \sin\delta_R
\cos(2\phi_B)$. Such a large difference can be detected at B factories.

The contribution from dimension 5 color dipole operator has been  
estimated by assuming that color octet operators 
contribute the amount as determined 
in generalized factorization approximation. The magnitude of the color
dipole coefficient as large as 10 times of the SM, if the 
chiral structure is the same as the SM one, is not ruled out and 
may in fact play some important role in the missing charm problem.
The weak phases $\sigma_{T,||,0}$ can be as large as $0.08\sin\delta_c$,
where $\delta_c$ is the weak phase of the color dipole phase~\cite{he4}.
The phases are approximately equal for $\sigma_T$ and $\sigma_*$. 
The phase $\sigma_S$ is suppressed by a factor of
$m_\psi^2/m_B^2$ due to helicity structure of the operator.
The value for $\Delta_{KK^*}$ can be as large as
$0.18\sin\delta_c \cos(2\phi_B)$.
If it turns out that the color dipole interaction has opposite chirality,
although there is an enhancement factor of $1/(1-2|P|^2)$, the coefficient 
$C_R$ is more stringently constrained resulting in a smaller $\Delta_{KK^*}$.

I now discuss direct CP violation in angular distribution for 
charge $B\to J/\psi K^*$. The full angular distribution for
$B\to J/\psi K^*$ is given by

\begin{eqnarray}
&&{1\over \Gamma} {d\Gamma \over d\cos\theta_{tr} d\cos \theta_{K^*}
d\phi_{tr}}= {9\over 32\pi} \{2|A_0|^2 \cos^2\theta_{K^*}(1-\sin^2\theta_{tr} \cos^2\phi_{tr})
\nonumber\\
&&+ |A_{||}|^2 \sin^2\theta_{K^*} (1-\sin^2\theta_{tr}\sin^2\phi_{tr})
+ |A_{T}|^2 \sin^2\theta_{K^*} \sin^2\phi_{tr}\nonumber\\
&&
- Im(A^*_{||} A_{T}) \sin^2\theta_{K^*} \sin2\theta_{tr}\sin \phi_{tr}\nonumber\\
&& + {1\over \sqrt{2}} Re(A^*_0 A_{||}) \sin2\theta_{K^*} \sin^2\theta_{tr}
\sin2\phi_{tr}\nonumber\\
&& +
{1\over \sqrt{2}} Im(A^*_0A_{T}) \sin2\theta_{K^*} \sin2\theta_{tr}\cos\phi_{tr}\}
,
\end{eqnarray}
where 
the transversity angles $\theta_{tr}$ and
$\phi_{tr}$ are defined as polar and azimuth angles of the charge lepton in the
$J/\psi$ rest frame with x axis along the direction of $K^*$, x-y plane 
parallel to $K\pi$ plane. The angle $\theta_{K^*}$ is defined as that of the 
$K$ in the rest frame of $K^*$ 
relative to the negative of the $J/\psi$ direction in that frame.

In the CLEO analysis, the FSI phase for 
$A_0$ was taken to be zero. For convenience
I will use the convention that each amplitude $A_i$ has both CP conserving
FSI phase $\phi_i$ and CP violating phase $\sigma_i$ as indicated before. 
With this convention, $A_j = |A_j| e^{i(\phi_j+\sigma_j)}$ while 
$\bar A_T = -|A_T|e^{i(\phi_T - \sigma_T)}$, 
$\bar A_{(||,0)} = |A_{(||,0)}|e^{i(\phi_{(||,0)} - \sigma_{(||,0)})}$.

It is clear that the coefficients $\alpha = - Im(A^*_{||} A_T)$, 
$\beta = Re(A^*_0 A_{||})$, and $\gamma = Im(A^*_0 A_T)$ of the last three
terms in the angular distribution, and similarly $\bar \alpha$, $\bar \beta$
and $\bar \gamma$ for $\bar B$ decays, contain information about CP violation.
Without separating $B$ and $\bar B$ decays, however, which was the case for
the CLEO analysis mentioned earlier, information on CP violation cannot be 
extracted. One must obtain the angular distributions for $B\to J/\psi K^*$ 
and $\bar B \to J/\psi \bar K^*$ decays separately and determine the coefficients
for the interference terms in each case. The following three quantities then 
measure CP violation~\cite{he5,gv}

\begin{eqnarray}
a_1 &=& \alpha + \bar \alpha 
=2|A_{||}||A_T|\cos(\phi_{||T})\sin(\sigma_{||T})
,\nonumber\\
a_2 &=& \beta - \bar \beta 
=-2|A_{||}||A_0|\sin(\phi_{||0})\sin(\sigma_{||0})
,\nonumber\\
a_3 &=& \gamma+\bar \gamma 
=-2|A_{T}||A_0|\cos(\phi_{T0})\sin(\sigma_{T0}),
\end{eqnarray}
where $\phi_{ij} = \phi_i-\phi_j$ and $\sigma_{ij} = \sigma_i -\sigma_j$.

It is interesting to note that the CP violating observable $a_{1,3}$ do not
require FSI phase differences and is especially sensitive to CP violating weak 
phases.
The present CLEO data on angular distributions
which provide information about FSI phases 
are the charge averaged CP conserving quantities which are proportional to
$\sin(\phi_{||T})\cos(\sigma_{||T})$,
$\cos(\phi_{||0})\cos(\sigma_{||0})$,
and $\sin(\phi_{T0})\cos(\sigma_{T0})$.
At present these data do not exclude 
$a_{1,3}$ up to 50\% and smaller for $a_2$

Again in the SM, the three CP violating observables are 
zero.  They provide good tests for CP 
violation beyond the SM there it is possible to have non-zero values for
$a_i$. For example in R-parity violating supersymmetric model considered
earlier, for
the central values for the magnitude of the amplitudes and assuming
that the FSI are zero, 
the asymmetries can be as large as 
$a_1 = 0.10\sin\delta_R$ and $a_3 = -0.12\sin\delta_R$.

The color dipole interaction to these asymmetries are small because with the
same chiral structure as  the SM, 
the new phases are approximately equal as discussed
before and therefore small asymmetries. For the case with opposite chiral 
structure as
the SM the strength of the interaction is more stringently constrained, the
phases are all small.

In all cases discussed above the weak phases for $\sigma_{||}$ and 
$\sigma_{0}$ are equal (or approximately equal).
The asymmetry $a_2$ is approximately zero in all cases considered, and
does not seem to be a good quantity to study for CP violation using this 
method.

The sensitivities for $a_{1,2}$ is similar to the 
sensitivity to the phase angles of the amplitudes. 
It is interesting to note that the systematic error in 
CLEO analysis is already as low as 0.03~\cite{Pwave}. 
With increases statistics, $a_{1,3}$ as large as 
0.10 should be accessible at CLEO, at CDF, and at 
future B factories. Since the errors are determined through a fit, it is not
clear how the statistical error scales with actual number of events. The 
question can only be answered by actual studies, but naive scaling implies that
one would need $10^8$ events to be able to distinguish the deviations given
earlier. Nevertheless, the needed number of events may be less and
measurement of the observables $a_i$ will provide us with useful information
about CP violation. 

I thank W.-S. Hou and G.-L. Lin for collaborations on the work reported here. 
This work is supported in part by 
ROC National Science Council and by Australian Research Council.

\end{document}